\documentclass[fleqn,10pt]{wlscirep}
\usepackage[utf8]{inputenc}
\usepackage[T1]{fontenc}
\title{Delay time of waves performing L\'evy walks in 1D random media}

\author[1,+]{L. A. Razo-L\'opez}
\author[1]{A. A. Fern\'andez-Mar\'in}
\author[1]{J. A. M\'endez-Berm\'udez}
\author[2]{J. S\'anchez-Dehesa}
\author[3,*]{V. A. Gopar}
\affil[1]{Instituto de F\'{\i}sica, Benem\'erita Universidad Aut\'onoma de Puebla,
Apartado Postal J-48, Puebla 72570, Mexico}
\affil[2]{Departamento de Ingenier\'ia Electr\'onica, Universitat Polit\`ecnica de Val\`encia, Camino de vera s.  n. (Edificio 7F), ES-46022, Valencia, Spain}
\affil[3]{Departamento de F\'isica Te\'orica, Facultad de Ciencias, and BIFI,  Universidad de Zaragoza, Pedro Cerbuna 12, ES-50009 Zaragoza, Spain}

\affil[+]{Present address: Universit\`e C\^ote d'Azur, CNRS, Institut de Physique de Nice, Parc Valrose. 06100 Nice, France}
\affil[*]{Email: gopar@unizar.es}


\begin{abstract}
The time that waves spend inside 1D random media with the possibility of performing L\'evy  walks is experimentally and theoretically studied. 
The dynamics of quantum and classical wave diffusion has been investigated in canonical disordered systems via the delay time. We show that 
a wide class of disorder--L\'evy disorder--leads to strong random fluctuations of the delay time; nevertheless, some statistical properties such as the tail of the  distribution and the average of the delay time are insensitive to L\'evy walks. Our results reveal a universal character of wave propagation that goes beyond standard Brownian wave-diffusion.
\end{abstract}

\begin{document}

\flushbottom
\maketitle
%
%
\thispagestyle{empty}

A wave packet launched into a scattering region can penetrate that region and it may be reflected eventually. Thus, one might wonder how much time the wave packet has spent inside the media.  
This fundamental question was addressed by Wigner and Smith  \cite{Wigner1955,Smith1960}. It was shown that the delay time $\tau_R$ of a wave packet is related to the derivative of the reflection phase 
$\theta_R$ with respect to the frequency $\omega$: $\tau_R = d\theta_R/d\omega$.

The delay time has received  attention in many  
disciplines since it reveals information on the scattering medium and, therefore, it has also been of interest from an application point of view; e.g.,
the delay time is a fundamental quantity in imaging of tissues in optical coherence tomography~\cite{Fercher_2003,Frank2020}.

A major issue in wave transport is the presence of disorder.
Moreover, if waves propagate coherently 
through 1D random media, complex interference effects emerge, such as the widely studied phenomenon of Anderson localization~\cite{Anderson1958,Lagendik2009}: an exponential decay in space of classical and  quantum waves,  for instance, electromagnetic waves and electrons, respectively.

Since disorder is ubiquitous in real systems, there has been a great interest in studying the effects of Anderson localization on dynamical quantities such as the delay time. 
Microwave experiments have been performed to analyze statistical properties of wave dynamics~\cite{Genack1999,Sebbah1999,Chabanov2001}, while several theoretical approaches have been developed to describe the delay-time  
statistics (see Ref.~\cite{Texier2016} for a review).

Remarkably, it has been demonstrated that some statistical properties of the delay time are invariant in the sense that they are independent of the details of the medium. For instance, the inverse square power decay of the distribution of $\tau_R$ has been predicted in semi-infinite 1D systems~\cite{Texier1999} and also studied in higher dimensions~\cite{Schomerus2000,Xu2011,Ossipov_2018}. Another example is  the 
average delay time, which is proportional to the mean length of trajectories 
\cite{Pierrat2014}. This quantity was predicted to 
be invariant  with respect to details of the scattering region, as 
recently observed experimentally \cite{Pierrat2014,Blanco2003,Savo2017}.

\begin{figure}
 \centering
\includegraphics[width=0.9\columnwidth]{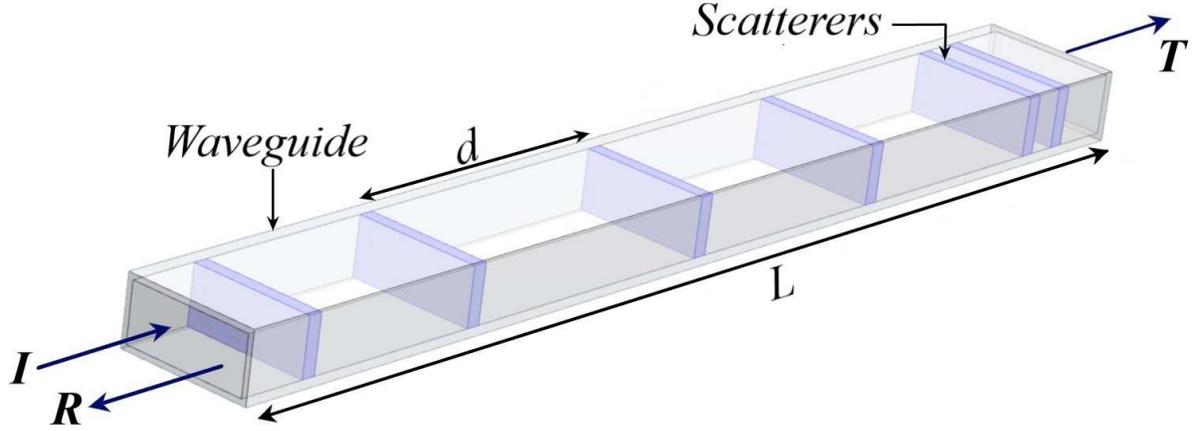}
\caption{(a) Schematic view of the experimental setup. The aluminum waveguide (1), containing randomly distributed dielectric slabs, is connected to the ports of a vector network analyzer (2) and the data is stored in a computer (3). (b) Actual 2m long aluminum waveguide (22.8mm width and 10.6mm height). The top is open to allow inner vision.
}
\label{Fig_0}
\end{figure}

Previous experimental and theoretical works on the delay time in 1D consider only models of disorder that lead to Anderson localization, however, there is a wide  class of disorder--L\'evy disorder--that leads to  delocalization or anomalous localization, in relation to the Anderson localization~\cite{Falceto2010,Ilias2012,Kleftogiannis2013,Barbosa2019}. Anomalous localization finds its origin in the  nonzero probability that waves travel a long distance without being scattered; these events are scarce but have a large impact \cite{flights}.

Here, we  experimentally and theoretically study the delay time of reflected microwaves suffering coherent multiple  scattering in a medium characterized by random spacings of scatterers following a one-sided L\'evy $\alpha$-stable distribution. 
Experiments in waveguides with  standard disorder (i.e., with random scatterer separations following a non-heavy tailed distribution) are also performed to compare results with those of L\'evy disorder. Additionally, numerical simulations are carried out to overcome some practical limitations of experiments and to obtain further support of our model. We calculate the distribution of the delay time and, furthermore, our results allow us to conclude that universal features of the delay time in canonical disordered media go beyond  standard Brownian models of wave diffusion, despite the fact that the presence of L\'evy  walks leads to stronger random fluctuations of the delay time.

L\'evy statistics has been found in a broad range  of contexts that 
go from foraging patterns of marine predators~\cite{Mantenga1995} to fluctuations of stock market indices~\cite{Reynolds2016} or resonant emission of light \cite{Patel2012}. L\'evy models are applied to describe  anomalous diffusion of particles and waves that cannot be described by standard Brownian models
\cite{Bouchad1990,Shlesinger1999,Klafter2000,Barthelemy2008,Solomon1993,Mercadier2009,Rocha2020,Zaburdaev2015}. 

Essentially, L\'evy random processes are characterized by probability distributions 
whose tails decay like a power-law, i.e., if $x$ is a random variable with probability density 
$\rho(x)$, then  $\rho(x) \sim 1/x^{1+\alpha}$ for $x \gg 1$ \cite{Uchaikin1999}. Fluctuations of random variables that follow L\'evy statistics are so large that the first and second moments diverge 
for $0 < \alpha <1$.

\section*{Results}
\subsection*{Experiments}

Microwaves are launched into 
an aluminum waveguide containing  2.5 mm thick dielectric slabs whose separations follow a L\'evy $\alpha$-stable distribution (see Fig. 1). We work in a frequency range where a single transport channel is supported.  
Two different $\alpha$- stable distributions characterized by their power-law decay
have been chosen: $\alpha=1/2$ and 3/4. Additionally, a conventional disordered microwave waveguide with random spacing between slabs following a Gaussian distribution has been built. Using a network vector analyzer, we measure   
the $2 \times 2$ scattering matrix $S$:
\begin{eqnarray}
\label{Smatrix}
S & = & \left(\begin{array}{cc}
\sqrt{R} e^{i\theta_R} & \sqrt{T} e^{i\theta_T} \\
\sqrt{T} e^{i\theta_T} & \sqrt{R}e^{i\theta_{R'}} \\
\end{array} \right) , 
\end{eqnarray}
where $R$ and $T$ are the reflection and transmission coefficients, respectively.  Measurements of $S$ are thus collected over different disorder realizations.

\begin{figure}
 \centering
\includegraphics[angle=-90,width=1\columnwidth]{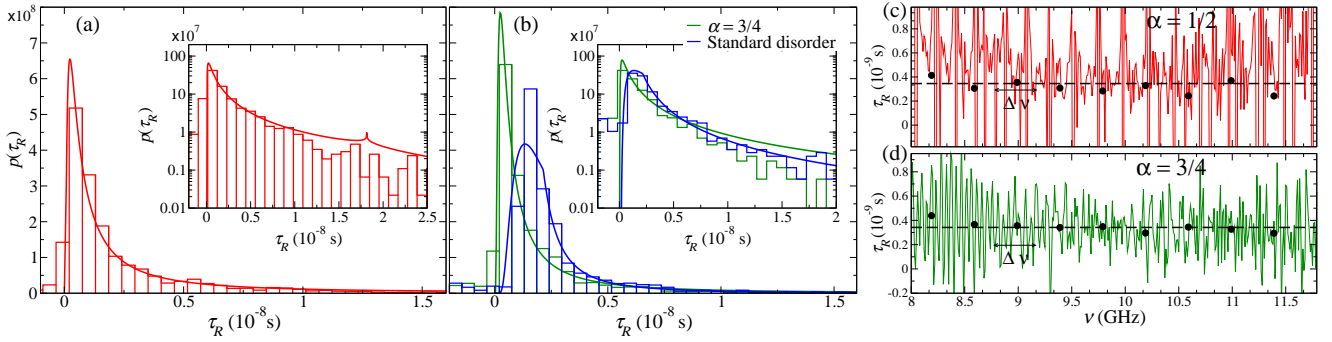}
\caption{Experimental delay-time distributions  (histograms) for L\'evy
waveguides characterized by
(a) $\alpha=1/2$ and $\left< -\ln T \right>=4.7$ at 9.9 GHz (red histogram) and
(b) $\alpha=3/4$ and $\left< -\ln T \right>=12$ at 11.2 GHz (green histogram).
The blue histogram in (b) corresponds to random waveguides with
ordinary (Gaussian) disorder with $\left< -\ln T \right>=12$ at
11.2 GHz.
The histograms were constructed with (a) 4590 and (b) 1890 data.
Insets show $p(\tau_R)$ in a logarithmic scale.
Red, green [blue] solid curves show the theoretical predictions from
Eq. (\ref{poftaualpha1}) [Eq. (\ref{poftaur_s})]. Experimental delay times for a typical realization of the disorder of waveguides with (c) $\alpha=1/2$ and (d) 3/4. Black dots represent the average of $\tau_R$ 
over frequency windows $\Delta \nu=0.4$GHz. The horizontal black dashed lines are the averages of $\tau_R$ over the complete frequency window (8-12 GHz).}
\label{Fig1}
\end{figure}

From the collected $S$-matrices, we obtain $\tau_R$ and the   probability distribution function   
$p(\tau_R)$. Figures \ref{Fig1}(a) and \ref{Fig1}(b)  show the distribution $p_\alpha(\tau_R)$ with 
$\alpha=1/2$ (red) and 3/4 (green), respectively. The insets show $p_\alpha(\tau_R)$ on a logarithmic scale for a better visualization of the tail. 
In Fig.~\ref{Fig1}(b), the delay-time  distribution (blue histogram) for  conventional disorder is also shown. 
Both distributions in  
Fig.~\ref{Fig1}(b) have the same average value $\langle \ln T \rangle$. 
We can observe that the profile of both distributions (green and blue histograms) is different.

The distributions $p_\alpha (\tau_R)$ from our model, which we introduce below,  are also shown (solid lines) in Figs.~\ref{Fig1}(a) and \ref{Fig1}(b). It is observed in Fig. \ref{Fig1}(a) 
that for $\alpha=1/2$, $p_\alpha(\tau_R)$ 
shows a small peak in the tail. The distribution for $\alpha=3/4$ in Fig. \ref{Fig1}(b) also exhibits  a peak but it is smoother and occurs at a larger value of $\tau_R$, outside of the time range shown in Fig. \ref{Fig1}(b). We attribute these peaks 
to scattering processes that reach  the 
right boundary of the waveguide; in L\'evy disordered samples those processes are favored since waves can travel long distances without being scattered. In contrast, for ordinary disordered systems,  $p(\tau_R)$ decays monotonically and for $\tau_R \gg 1$, $p(\tau_R) \sim 1/\tau_R^{2}$~\cite{Texier1999}.

The trend of the experimental distributions (histograms) is well described by the model (solid lines), despite the fact that the statistics 
of $\tau_R$ is extracted from a limited amount of experimental data
and the presence of a small tail for  negative values of $\tau_R$ observed in  
Figs. \ref{Fig1}(a) and \ref{Fig1}(b). Negative delay times are not considered in our model and are thus a source for discrepancies between experimental and theoretical results. It has been proposed that those negative values are due to a strong distortion of the wave packet produced due to interference between incident and promptly reflected waves~\cite{Garrett1970,Dalitz1979,Chu1982,Durand2019}.

We now address an invariance property 
of the mean path of trajectories with respect to the details of the disordered medium. This invariance property is equivalent to the independence  
of the average delay time with energy since both quantities are proportional~\cite{Pierrat2014}. To illustrate this invariance,  in Figs. \ref{Fig1}(c) and \ref{Fig1}(d), $\tau_R$  is plotted as a function of the linear frequency $\nu$ for typical samples with  $\alpha=1/2$ and 3/4,  respectively. We see  strong fluctuations of $\tau_R$, however, the average delay-time  over frequency windows $\Delta \nu(=0.4$GHz) is essentially independent of the frequency, as it is observed in both figures (dots).
Moreover, the average over the whole frequency window (horizontal dashed line) has  the same value ($3.4\times 10^{-9}$s) for both cases: $\alpha=1/2$ and 3/4, and thus, it is independent of particularities of 
the medium. We will address later this point in more detail.

\subsection*{Model}
For conventional disorder and 
within a random matrix approach to localization~\cite{Anderson1980,Mello2004}, the mean free path $\ell$ determines  the statistical properties of  the transport  
and it is related to the transmission by  $s \equiv \langle -\ln T \rangle=L/\ell$, which  is  proportional to the number of scatterers $n$ in the system, i.e., 
$\langle -\ln T \rangle = b n$ with $b$  constant~\cite{Mello1986}. If the random spacing between scatterers follows a L\'evy $\alpha$-stable distribution, the number of scatterers in a system of length $L$ is subject to strong random fluctuations. Such fluctuations are described by the probability density $\Pi_L(n;\alpha)$ given by \cite{Falceto2010}: $\Pi_L(n;\alpha)=2L q_{\alpha,1}\left(L/(2n)^{1/\alpha}\right)  /\alpha (2n)^{(1+\alpha)/\alpha}$, where $q_{\alpha,c}(x)$ is the probability density function 
of the L\'evy $\alpha$-stable distribution with exponent $\alpha$ and scale parameter $c$. For $x \gg 1$, $q_{\alpha,c}(x) \sim c/x^{1+\alpha}$. 
Therefore, with the knowledge of $\Pi_L(n;\alpha)$ and the delay-time probability density for canonical disordered systems, $p_s (\tau_R)$, we write the probability density $p_\alpha (\tau_R)$ for L\'evy disordered systems as 
\begin{equation}
\label{poftaualpha}
p_\alpha(\tau_R)=\int_0^{\infty}p_s(\tau_R) \Pi_L(n;\alpha) dn. 
\end{equation}
The  probability density $p_s(\tau_R)$ in its full generality remains, however, an open problem. For semi-infinite disordered systems,  assuming no transmission, the limit ($L \to \infty$)  
delay-time distribution $p_{\infty}(\tau_R)$ is given by    \cite{Jayannavar1989,Heinrichs1990,Comtet1997,Texier1999,Bolton-Heaton1999,Beenakker2001}:  
$ p_{\infty}(\tau_R) = \tau_\ell \tau_R^{-2} \exp{(-\tau_\ell/\tau_R)}$, 
where $\tau_\ell$ is the scattering time of the disorder. Real systems, however, are finite and finite-size effects may be of relevance. In particular, our experiments are performed in 2 m long waveguides and microwaves can be transmitted. 

Our model for $p_s(\tau_R)$ that will be verified experimentally and numerically
involves two main assumptions. Firstly, though
the waves in our waveguides can be transmitted, we use 
a relationship between $\tau_R$ in the absence of absorption and $R$ in the presence of weak absorption that assumes negligible transmission: $R=1-\tau_R/\tau_0$, where $\tau_0$ is the absorption time  which is assumed  very large~\cite{Klyatskin1992,Ramakrishna2000,Beenakker2001}. A key point is that this relationship establishes that fluctuations of $R$ determine the statistics of $\tau_R$. Notice that for an ensemble of different samples, $\tau_0$  fluctuates since  it depends on the disorder configuration. Indeed,  later $\tau_0$ will be identified with $\langle \tau_R \rangle$, which contains information of the system length. Secondly, 
we use the Laguerre ensemble $p(\mu)$ which describes the statistics of $R$ assuming  large samples~\cite{Beenakker1996,verification}. That is, $p(\mu) \propto \exp{\left(- \gamma \mu \right)}$ where  $\mu^{-1} \equiv R-1$ and $\gamma$ is a constant. For systems with absorption  ($R < 1$) 
$\gamma < 0$ and $\mu < -1$, while for systems with amplification ($R>1$), $\gamma > 0$ with  $\mu >0$. For the latter case $R=1+\tau_R/\tau_0$. Let us stress that 
for systems of finite length,  $\tau_R$  can exceed $\tau_0$ since, for instance, $\tau_R$ is infinite for transmitted waves. We thus need to 
consider both cases $\tau_R < \tau_0$ and $\tau_R > \tau_0$. 
Therefore, after the change of variable $\mu \to \tau_R$, the normalized distribution  $p_s(\tau_R)$ can be expressed as

\begin{figure}
\centerline{\includegraphics[width=0.5\columnwidth]{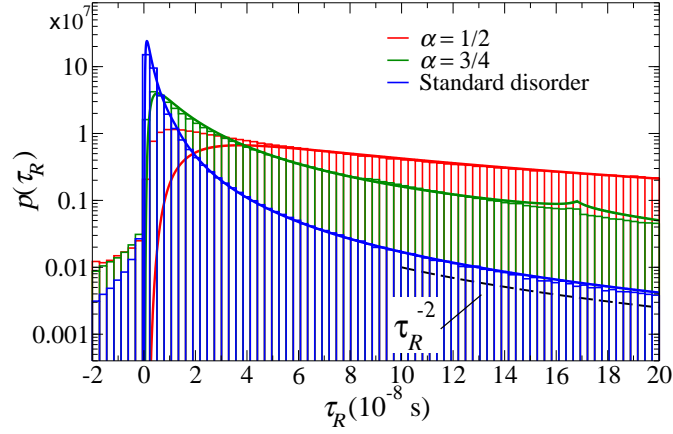}}
\caption{Numerical delay-time  distributions with  parameters $\alpha=1/2$ (red histogram) and $\alpha=3/4$ (green histogram) and for random waveguides with conventional (Gaussian) disorder (blue histogram). In all cases
$\left< -\ln T \right> = 10$.
The histograms were constructed with $5 \times 10^6$ disorder realizations.
The solid curves are the theoretical predictions from Eqs. (\ref{poftaur_s}) and (\ref{poftaualpha1}) 
for standard and L\'evy disorder, respectively.
The dashed line, proportional to $1/\tau_R^{2}$, is a guide to the eye.}
\label{Fig2}
\end{figure}
\begin{equation}
\label{poftaur_s}
 p_s(\tau_R) = \frac{a}{s \left[2-\exp{(-a/s)}\right]}\frac{1}{\tau_R^2} 
 \exp\left( {-a \left| 1/\tau_R -1\right|/s} \right),
\end{equation}
where  $a\equiv 2L/(v_g \tau_0)$ and $v_g$ is the group velocity. In writing Eq.~(\ref{poftaur_s}), we 
conveniently measure the delay time in units of 
$\tau_0$, i.e., we replaced $\tau_R/\tau_0 \to \tau_R$. The above expression for $p_s(\tau_R)$ is obtained after making $\mu \to |\mu+1|$ and the change of variable $\mu \to \tau_R$ in the Laguerre ensemble $p(\mu)$  with $-\mu^{-1}=1-R=\tau_R/\tau_0$.
Notice that with the absolute value in Eq. (\ref{poftaur_s}), both cases $\tau_R/\tau_0<1$ and $\tau_R/\tau_0>1$ are considered. 
Our assumptions may overestimate 
$p_s(\tau_R)$,  mainly  for $\tau_R > \tau_0$, since some of the scattering processes that reach the right end of the sample may leave the waveguide, 
having actually an infinite reflection delay time. Thus, as the systems become shorter, discrepancies between our model and experiments or simulations are expected. On the other hand, Eq. (\ref{poftaur_s}) reduces to 
$p_{\infty}(\tau_R)$ for  $\tau_0/\tau_R \gg 1$. Additionally, since Eq. (\ref{poftaur_s}) is based on the Laguerre ensemble, in the SM we have 
experimentally and numerically verified the distribution of the reflection coefficient predicted by the Laguerre ensemble.

We now define $s(z,\alpha,\xi) \equiv \xi/(2 z^\alpha f(\alpha))$ with 
$z=L/(2 n)^{1/\alpha}$ and, for a system of fixed $L$, $\xi=\langle -\ln T \rangle_L$ which is  given by 
\cite{Falceto2010}: $ \langle - \ln T \rangle_L=b L^\alpha f(\alpha)/c$, where $f(\alpha)=(1/2)\int_0^\infty z^{-\alpha} q_{\alpha,1}(z)dz$.  
Therefore, using Eqs. (\ref{poftaualpha}) and (\ref{poftaur_s}), we  write the 
distribution of $\tau_R$ for L\'evy disordered samples as 
\begin{equation}
\label{poftaualpha1}
 p_{\alpha}(\tau_R)= \int_0^\infty p_{s(z, \alpha,\xi)}(\tau_R) q_{\alpha,1}(z) dz ,
\end{equation}
where $p_{s(z, \alpha,\xi)}(\tau_R)$ is given in Eq. (\ref{poftaur_s}) with $s$ replaced by  $s(z,\alpha,\xi)$. 
The experimental distributions in Fig. \ref{Fig1} have been compared 
with Eqs. (\ref{poftaur_s}, \ref{poftaualpha1}) using $\tau_0$ as a fitting parameter.

Numerical simulations are now performed  for further support of Eq.~(\ref{poftaualpha1}) and to reveal universal properties. Thus, the number  
of disorder realizations is greatly increased  and absorption, which reduce the effects of long trajectories, is absent. Details of the numerical simulations are provided  in the SM.

Figure  \ref{Fig2} compares  numerical and theoretical delay time distributions from Eqs.~(\ref{poftaur_s}) and (\ref{poftaualpha1}) for standard and  L\'evy disordered waveguides with $\alpha=1/2$ and 3/4 and  $\xi=10$. For conventional disorder, our simulations show a physically meaningful result: the absorption time $\tau_0$ can be identified with the average $\langle \tau_R \rangle$. 
After this identification, since  $\langle \tau_R \rangle$ can be extracted from the numerical simulations, there is no free fitting parameters in Eq. (\ref{poftaur_s}). Similarly, 
for L\'evy disorder $\tau_0$ has been found to fit the numerical distributions with $\tau_0=2\langle \tau_R \rangle/\alpha$ (for standard disorder, $\alpha=2$). This result is appealing, however, its formal demonstration remains as an open problem.

Although the distribution profiles in Fig. \ref{Fig2} are different and show the impact of L\'evy walks, they share some properties that are not evident  because of the different time scale of each case.

\begin{figure}
\centerline{\includegraphics[width=0.7\columnwidth]{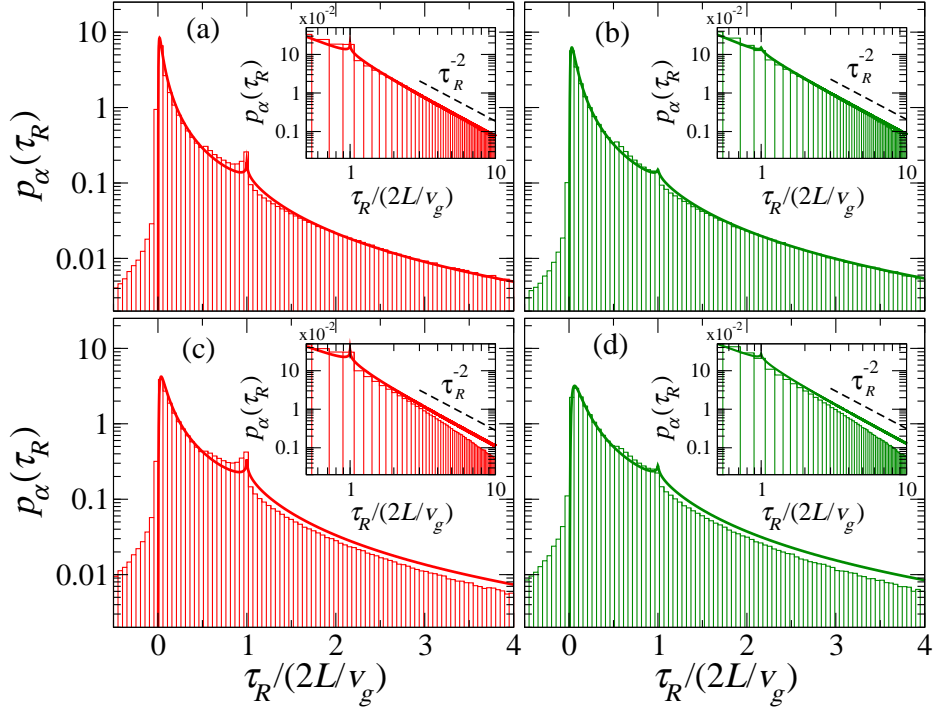}}
\caption{Numerical distributions $p_\alpha(\tau_R)$ (histograms) for L\'evy disordered systems 
characterized by (a,c) $\alpha=1/2$ and (b,d) $\alpha=3/4$
with (a,b) $\xi=10$ and (c,d) $\xi=4$. Each histogram was obtained from $5\times 10^6$ disorder realizations. The solid curves are the theoretical predictions from Eq. (\ref{poftaualpha1}).
Insets show $p_\alpha(\tau_R)$ in a logarithmic scale. 
The dashed lines, following the power law $1/\tau_R^{2}$, are a guide to the eye (see also  SM).}
\label{Fig3}
\end{figure}

In order to compare $p_{\alpha}(\tau_R)$ for different system parameters, it is convenient to express $\tau_R$ in units of $2L/v_g$, as shown in  Fig.~\ref{Fig3}  for $\alpha=1/2$ and 3/4, left and right panels, respectively, and  $\xi=10$ and 4, upper and lower panels,  respectively.  
A notorious difference with respect to the distribution for standard disordered systems (Fig.~\ref{Fig2}, blue line) is that a peak appears at $\tau_R/(2L/v_g)=1$, which is precisely the time that a wave would spend on traveling back and forth between the boundaries of the waveguide in the absence of disorder.

For shorter systems [Figs. \ref{Fig3}(c) and \ref{Fig3}(d)], we notice a deviation, mainly at the distribution tails, of the theoretical predictions (solid lines) with respect to the numerical results. Also, the numerical simulations start to deviate from the $1/\tau_R^2$ decay [see insets in Figs. \ref{Fig3}(c) and \ref{Fig3}(d)], which is expected since the transmission is higher as the waveguide becomes shorter.

We now show  that some properties of $\tau_R$ go beyond canonical disorder models. 
We have already mentioned the inverse square decay of the delay time  distributions for $\tau_R \gg 1$ obtained in 1D  semi-infinite Anderson localized systems;  indeed, such power-law behavior has been explained by resonance models in the localized regime \cite{Texier1999,Bolton-Heaton1999,Kottos2005}. For L\'evy disordered structures, the insets of Figs. \ref{Fig3}(a) and \ref{Fig3}(b) compare the tail of the distributions  for 
$\alpha=1/2$ and 3/4, respectively, with  $1/\tau_R^{2}$ decay (dashed lines); 
see also the SM for further details and numerical fits of the tails. 
Actually, from Eqs. (\ref{poftaur_s}) and (\ref{poftaualpha1}), we also find that $p_\alpha(\tau_R) \sim 1/\tau_R^{2}$ for $\tau_R\gg1$. 

\begin{figure}
\centerline{\includegraphics[width=0.7\columnwidth]{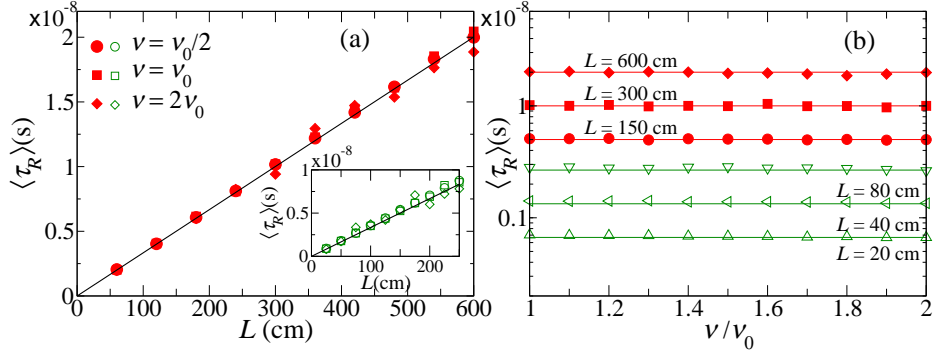}}
\caption{(a) Numerical average delay-time  $\langle \tau_R \rangle$ for L\'evy disorder 
 as a function of the length $L$ for different linear frequencies $\nu$ ($\nu_0=7.5$ GHz) with  $\alpha=1/2$ (main frame) and 
 $\alpha=3/4$ (inset). (b) $\langle \tau_R \rangle$  as a function $\nu$ for several lengths $L$ for 
$\alpha=1/2$ (red symbols) and $\alpha=3/4$ (green  symbols).
The averages are obtained from $10^5$ disorder realizations.
The horizontal solid lines correspond to $\left< \tau_R \right> = L/v_g$ in each case.}
\label{Fig4}
\end{figure}

Additionally, the average  $\langle \tau_R \rangle$ is a linear function of the system length: $\langle \tau_R \rangle = L/v_g$, as shown in Fig.~\ref{Fig4}(a) for $\alpha=1/2$ and 3/4 (inset), which is also observed in standard disordered systems \cite{Comtet1997,Texier1999}. Furthermore, an interesting invariance property of the  mean length of random walk trajectories with respect to the details of the disorder~\cite{Blanco2003} has been recently investigated in optical experiments~\cite{Savo2017}. The invariance of the mean path length  is equivalent to the independence of the average delay-time  to the energy. We have already shown experimental evidence of this invariance in Figs. \ref{Fig1}(c) and \ref{Fig1}(d).  Figure~\ref{Fig4}(b) provides additional  
numerical evidence by showing  $\langle \tau_R \rangle$ 
for disordered systems of different lengths with  $\alpha =1/2$ and 3/4. 
We observe that $\langle \tau_R \rangle$ is constant with the linear frequency  $\nu$. Thus, these results give further evidence that the invariance of the  mean path length goes beyond Brownian random walk models. This invariance can be explained by a direct relation between the delay time and the density of states \cite{Smith1960}, as was studied in \cite{Pierrat2014} from ballistic to localized regimes.

\section*{Discussion}

We have studied experimentally and theoretically the impact of L\'evy walks on a fundamental dynamical quantity: the delay time. Although the delay-time distributions for L\'evy 
and canonical disordered structures are  different, remarkably, some  properties are invariant, e.g., the inverse square power-law decay of the delay-time distribution and the insensitivity of the average delay time to energy. The latter is equivalent 
to the invariance of the mean path length observed in experiments of light propagation. Additionally, the linear dependence of the average delay time  with the length in ordinary disordered media is not affected by the presence of L\'evy walks.  All together, our results  reveal a universal character of wave propagation that goes beyond standard Brownian models \cite{Pierrat2014,Blanco2003,Savo2017}.

We  point out that in L\'evy disordered systems 
with $\alpha<1$, the mean free path is meaningless since it diverges, in contrast to canonical disordered systems in which the mean free path settles the wave statistics.  
In the presence of L\'evy type of disorder, two quantities determine the transport statistics: $\alpha$  and the logarithmic transmission average.

Our model shows a good agreement with experimental and numerical results, however, only structures with a single transport channel or 1D systems have been considered.  Although nowadays 1D wave transport is of relevance, it would be desirable to extend our study to higher dimensions. Nevertheless, the  ideas and results presented here are so general that can be applied from classical to quantum waves in disordered media.

\section*{Methods}

To determinate experimentally the delay times $\tau_R$, we employ the experimental setup schematically described in Fig.~\ref{Fig_0}. We use a 2 meters long aluminum waveguide with a rectangular cross-section (22.8 mm width and 10.6 mm height) in which we insert a random distribution of dielectric scatterers. Each scatterer consists of a 2.5 mm thick dielectric slab.  The slabs are made of FR4, a composite material with low losses, its dielectric permittivity is $\epsilon=4.4+i0.088$. 

With those dimensions of the waveguide, 
the fundamental mode TE10 propagates along the waveguide in the frequency range from 7.5 to 15 GHz. Within this frequency window, the waveguide can be effectively considered as a 1D random waveguide. 
A two-port vector network analyzer (VNA) ZVA 24 from Rohde \& Schwarz (2 in Fig.~\ref{Fig_0}) with  a resolution of 1 Hz is employed to generate and receive microwave signals in the region of interest. The ports of the VNA are used alternatively as sources and receivers of the microwave signals. Each port is connected to the aluminum waveguide by means of standard X-band microwave transitions placed at both sides of the waveguide. This allows the measurements of the four matrix elements of the complex $S$-matrix, $S(\nu)$, where $\nu$ is the linear frequency. In particular, the phase of the reflection amplitude $\theta_R(\nu)$ is obtained as given in Eq.~(1).

From the set of measured phases $\theta_R(\nu_i)$, we compute the delay times 
$\tau_R(\nu) = d\theta_R(\nu)/d\omega$ as 
\[
\tau_R(\nu_i) \approx \frac{1}{2\pi}\frac{\theta_R(\nu_{i+1})-\theta_R(\nu_i)}{\Delta \nu}, 
\]where $\Delta\nu=(\nu_{i+1}-\nu_{i})=0.01$GHz.

We have built waveguides with both L\'evy and canonical types of disorder. 
For the L\'evy waveguides, the dielectric slabs have been placed accordingly to a $\alpha$- stable distribution with parameters $\alpha=1/2$ and 3/4, i.e., $\rho(d)\sim 1/d^{1+\alpha}$ for $d\gg 1$. Let us comment that in general there are no closed-form expressions in terms of elementary functions for the L\'evy $\alpha$-stable distributions\cite{fourier,Calvo2010} (except for $\alpha =1/2$, 1 and 2). For any  value of $\alpha$ with $0< \alpha <2$, the 
L\'evy $\alpha$-stable distribution is  generated numerically~\cite{Uchaikin1999}.

For the canonical disorder, the separation between slabs follows a standard  normal distribution (zero mean and unit variance).

An ensemble of  135 random waveguides of different disorder realizations have been built for each case: $\alpha=1/2$, 3/4, and standard disorder. We  measured the reflection amplitude in a frequency window centered at 
$\nu=9.9$GHz for L\'evy disordered waveguides with $\alpha=1/2$. For 
$\alpha=3/4$ and for conventional disorder, the reflection amplitude was measured in a frequency window centered at $\nu=11.2$GHz. Thus, the experimental histograms of Figs.~\ref{Fig1}(a) and~\ref{Fig1}(b)  have been obtained from 4590 and 1890 delay times, respectively.


\begin{thebibliography}{99}
\bibitem{flights}
Such kind of events are sometimes called L\'evy flights, however, 
L\'evy flights is mostly used to study processes where the jumps or flights are instantaneous, while in L\'evy walks, the velocity of the flights is constant, as in our study.
\bibitem{fourier} An explicit closed  expression for $q_{\alpha,c}(x)$ can be conveniently given using the Fourier transform \cite{Calvo2010}:
$\bar q_{\alpha,c}(k) =\exp{\left[-|k|^\alpha \left(B H(k)+\overline{B} H(-k) \right)  \right]}$, where  H is the Heaviside step function and $\overline{B}$ is the complex conjugate of 
$B=-c \Gamma(-\alpha)\exp{\left( i\pi \alpha/2 \right)}$, $\Gamma$ being the Gamma function.
\end{thebibliography}


\section*{Acknowledgments}

A. A. F.-M. thanks the hospitality of the Laboratoire d'Acoustique de l'Universit\'e du Mans, France, where part  of this work was done. J. A. M.-B, gratefully acknowledges to Departamento de Matem\'{a}tica Aplicada e Estat\'{i}stica, Instituto de Ci\^{e}ncias Matem\'{a}ticas e de Computa\c{c}\~{a}o, Universidade de S\~{a}o Paulo during which this work was completed. 
J.A.M.-B. was supported by  
FAPESP (Grant No.~2019/06931-2), Brazil.
A. A. F.-M. thanks partial support by RFI Le Mans Acoustique and by the project HYPERMETA funded under the program Étoiles Montantes of the Region Pays de la Loire. V. A. G. acknowledges support by MCIU (Spain) under the Project number PGC2018-094684-B-C22.

\section*{Author contributions statement}

L. A. R-L performed the numerical simulations. A. A. F-M.  
and J. S-D. conducted and designed the experiments. 
L. A. R-L, J. A. M-B, and V. A. G. analyzed the results and contributed to the preparation of the manuscript. All authors reviewed 
the manuscript.

\section*{Additional information}

\textbf{Competing interests:} 
The authors declare no competing interests.

\end{document}